\documentclass[twocolumn]{aastex631}

\usepackage{xcolor}
\usepackage{graphicx}
\usepackage{epstopdf}
\usepackage[utf8]{inputenc}
\usepackage[T1]{fontenc}
\usepackage{url}
\usepackage{amsmath} 
\usepackage{enumitem}
\usepackage{booktabs}
\usepackage{natbib}
\usepackage{xspace}

\newcommand{\ha}{H$\alpha$\xspace}
\newcommand{\hb}{H$\beta$\xspace}
\newcommand{\oii}{[O\,{\sc ii}]\xspace}
\newcommand{\oiii}{[O\,{\sc iii}]\xspace}
\newcommand{\mgii}{Mg\,{\sc ii}\xspace}
\newcommand{\cgm}{CGM\xspace}

\newcommand{\desi}{DESI\xspace}

\newcommand{\kms}{km\,s$^{-1}$\xspace}

\newcommand{\Ha}{\hbox{{\rm H}$\alpha$}}

\newcommand{\Ovi}{\hbox{{\rm O}\kern 0.1em{\sc vi}}}
\newcommand{\OIII}{\hbox{[{\rm O}\kern 0.1em{\sc iii}]}}
\newcommand{\OII}{\hbox{[{\rm O}\kern 0.1em{\sc ii}]}}

\newcommand{\NII}{\hbox{[{\rm N}\kern 0.1em{\sc ii}]}}
\newcommand{\SII}{\hbox{[{\rm S}\kern 0.1em{\sc ii}]}}

\newcommand{\sigall}{<2}               
\newcommand{\sigjet}{5}              
\newcommand{\thetajet}{20^\circ}       
\newcommand{\fha}{1.19}                 

\begin{document}

\def \spose#1{\hbox  to 0pt{#1\hss}}  



\title{Lighting Up the CGM: Strong, Jet-Aligned $H\alpha$ Emission around Radio Galaxies}

\shorttitle{H-alpha CGM stacking }

\shortauthors{Roy et al.}


\correspondingauthor{Namrata Roy}
\email{namratar@asu.edu}

\author[0000-0002-4430-8846]{Namrata Roy}
\affiliation{School of Earth and Space Exploration, Arizona State University, Tempe, AZ 85287}

\author[0000-0002-2724-8298]{Sanchayeeta Borthakur}
\affiliation{School of Earth and Space Exploration, Arizona State University, Tempe, AZ 85287}

\author[0000-0001-6670-6370]{Timothy Heckman}
\affiliation{The William H. Miller III Department of Physics and Astronomy, The Johns Hopkins University, Baltimore, MD 21218, USA}
\affiliation{School of Earth and Space Exploration, Arizona State University, Tempe, AZ 85287}

\author[0009-0001-5959-9105]{Tanmay Singh}
\affiliation{School of Earth and Space Exploration, Arizona State University, Tempe, AZ 85287}


\begin{abstract}
A primary question within galaxy evolution is how active galactic nuclei (AGN) feedback modifies the circumgalactic medium (CGM). We present a search for faint H$\alpha$ emission from the cool ionized CGM ($T\sim 10^4$ K) around radio galaxies by stacking background-quasar spectra from DESI sightlines. We take into account the projected distance and position angle of each quasar sightline relative to the radio jet axis, and test whether jet--CGM coupling is anisotropic.  We detect a strong H$\alpha$ excess at $>5\sigma$ along the collimated radio jet axis ($\theta<20^\circ$) with a mean integrated flux of $1.19\times10^{-17}\ {\rm erg\ cm^{-2}\ s^{-1}}$. In contrast, the azimuthally averaged stack over all 324 sightline angles yields no detection ($<2\sigma$), indicating that this excess emission is very localized along the radio jet. We also find that the jet-aligned \ha\ signal is radially structured, where the strongest emission occurs near the host galaxy just outside the optical half-light radius, and rising again near the projected radio-lobe region. The jet-aligned stacks reveal \Ha \ signal that is roughly 100 times brighter than normal halos. In the same sightlines however, Mg\,{\sc ii} absorption shows no difference in incidence between jet-aligned and off-axis directions, with broadly similar equivalent widths, column densities, and line widths. This striking contrast shows that while Mg\,{\sc ii} traces the ambient, clumpy cool CGM reservoir, the H$\alpha$ emission directly captures localized, low-covering-fraction clouds whose density, pressure, or ionization level has been dramatically boosted by the propagating jet. These results deliver clear evidence of localized jet-CGM interaction in radio-jetted AGNs.

\end{abstract}



\section{Introduction} \label{sec:Introduction}

The circumgalactic medium (CGM) is the baryon reservoir that regulates the long-term growth of galaxies. It supplies gas for future star formation, receives recycled and outflowing material, and mediates the exchange of mass, metals, momentum, and energy between galaxies and the intergalactic medium \citep[e.g.,][]{Tumlinson2013,Borthakur2015,Borthakur2016,Tumlinson2017,Donahue2022}. A full accounting of these baryons, and understanding how they are heated, cooled, ejected, or reaccreted, remains a longstanding goal of galaxy evolution studies. 
This is especially important for massive quenched galaxies, where cool halo gas can survive long after active star formation has shut down \citep{Thom2012}. Since massive galaxies maintain quiescence over gigayear timescales,
a long-lived heating source is necessary to prevent that gas from cooling too efficiently and preventing the rejuvenation of star formation \citep{Cowie1977, Benson2003, Bower2006}.

There is now significant evidence that feedback from active galactic nuclei (AGNs) plays a crucial role in this regard, and  relativistic radio jets from the central supermassive black holes provide one of the clearest mechanical channels through which black holes can deliver energy beyond the nucleus.  \citep{Fabian1977, Ho2008, fabian12, croton06, mcnamara12, Kondapally2023}.   In galaxy clusters, radio jets inflate cavities, drive shocks, and redistribute energy through the hot X-ray-emitting atmosphere, demonstrating that radio-mode feedback can couple to halo gas on large scales \citep[e.g.,][]{Boehringer1993,Fabian2003,McNamara2005,Birzan2008,Gitti2012,fabian12,mcnamara12,heckman14,Hardcastle2020}. However, this direct evidence primarily traces hot gas with $T\gtrsim10^6$ K in massive group and cluster environments. How jet energy is deposited into the cooler, $T\sim10^4$ K CGM around the more common population of lower mass radio galaxies (log $M_{\star} < 11.5$) is much less understood. 

Existing observations suggest that jets can indeed kinematically affect the cool gas in the interstellar medium (ISM) and inner ISM--CGM boundary. In powerful radio galaxies at $z>2$, optical and near-infrared integral-field spectroscopy revealed broad, turbulent ionized gas components that are spatially aligned with the radio lobes, with line widths reaching ${\rm FWHM}\sim1000~{\rm km~s^{-1}}$  on kpc to tens-of-kpc scales \citep[e.g.,][]{nesvadba06,nesvadba08,nesvadba17,roy24,roy25,saxena24}. At larger radii, extended Ly$\alpha$ and H$\alpha$ halos have been observed in a handful of high redshift jetted AGNs sitting in massive halos at spatial scales often exceeding $100$ kpc from the nucleus \citep[e.g.,][]{McCarthy1993,van97,VillarMartin2003,Miley2008}. These systems show that cool and warm gas can coexist with powerful radio activity, but the prevalence of such jet-CGM coupling in more typical lower-redshift radio galaxies remains largely unconstrained because the CGM emission is extremely faint. Although the mechanical energy budget of radio jets is sufficient to affect the host galaxy  \citep[e.g.,][]{Bicknell1995,McNamara2005,Birzan2008,mcnamara12, roy21c, roy24, roy25}, where and how this energy is deposited into the cool CGM remains an open question. 

In this letter, we overcome this limitation by stacking optical spectra of background quasars from the large spectroscopic dataset obtained by the Dark Energy Spectroscopic Instrument (DESI) survey \citep{DESI2024, DESI2026}. We identify background quasar sightlines that pass through the projected halos of foreground radio galaxies from LOFAR Two-metre Sky Survey \citep[LoTSS;][]{Shimwell2017,Shimwell2022,Hardcastle2023}. This sample allows us to search for faint  emission and absorption CGM tracers in sightlines close to the large-scale radio jets.
Using the statistical power of large sample, we can detect extremely faint hydrogen recombination emission (H$\alpha$) from cool ionized gas along with magnesium resonance doublet absorption (Mg II) in galaxy halos, extending the measurements well into the CGM at $> 200$ kpc spatial distances \citep{Zhang2016, Chang2024}.

Our technique to isolate sightlines close to the radio axis is crucial to derive the impact of jets on the ambient medium, and significantly differs from previous studies \citep{Chang2024}. 
If radio jets couple to the cool CGM anisotropically, then the emission features should be enhanced preferentially along the radio lobes, while an azimuthally averaged stack may show little or no excess. This provides a direct statistical test of whether the CGM around radio galaxies behaves as a passive reservoir of gas, or as a medium that records where mechanical feedback energy has been deposited.

Throughout this Letter, we adopt a flat $\Lambda$CDM cosmology with $H_0=70$ \kms\ Mpc$^{-1}$, $\Omega_{\rm m}=0.3$, and $\Omega_\Lambda=0.7$. Projected distances are physical unless otherwise stated.

\section{Sample and stacking method}
\label{sec:sample_stacking}

\subsection{Data: DESI Quasar survey and Radio catalogs}
\label{subsec:data}

We construct a  foreground--background sample by combining radio galaxies by crossmatching spectroscopically confirmed background quasars from the DESI Data Release 1 \citep{Chaussidon2023, DESI2024, DESI2026} with foreground radio galaxies from the LOFAR Two-metre Sky Survey (LoTSS) DR2 value-added AGN catalog \citep{Shimwell2022, Hardcastle2023}. 
 DESI is a highly multiplexed optical spectroscopic survey on the Mayall 4 m telescope, with 5000 fibers and a wavelength coverage of approximately $3600$--$9800$ \AA\ at spectral resolution $R\sim2000$--$5000$ \citep{DESI2016,Silber2023,Guy2023}. The broad wavelength coverage makes DESI optimally suited for searching faint foreground \Ha \  emission and \mgii absorption in the background quasar spectra in and around the gaseous halos of radio galaxies. 
For each background sightline, we extract the final recommended quasar redshift from the value-added catalog $z_{\rm QSO}$, observed-frame wavelength, flux-density, and inverse-variance arrays. 

LoTSS DR2 provides $120$--$168$ MHz imaging over $27\%$ of the northern sky, with continuum maps at $6''$ resolution and a central frequency of $144$ MHz. The DR2 source catalog contains $4{,}396{,}228$ radio sources, with a median rms sensitivity of $\simeq83~\mu{\rm Jy~beam^{-1}}$ \citep{Shimwell2022}. 
We use the LoTss radio-optical value added catalogs to obtain their optical counterparts, redshifts, radio luminosities, angular sizes, and radio position angles \citep{Hardcastle2023}. We retain only sources with secure spectroscopic redshifts, and excluded objects where only photometric redshift estimates are available, because the foreground galaxy redshift sets the rest-frame wavelength grid for the H$\alpha$ stack and determines the projected physical separation of each quasar sightline. 


\begin{figure}[t]
    \centering
    \includegraphics[width=0.5\textwidth]{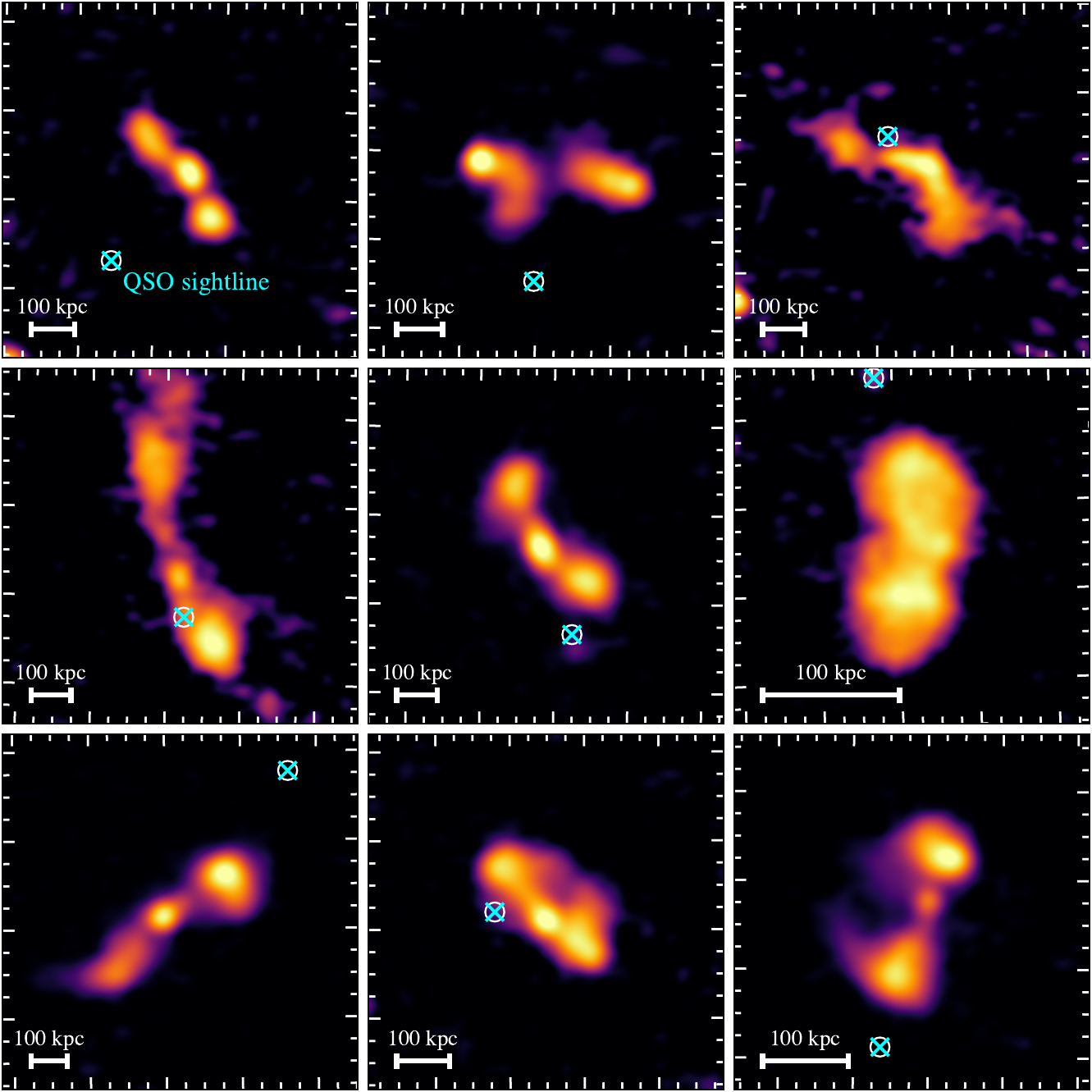}
    \caption{LOFAR radio images of 9 example radio galaxies used in our analyses. The background quasar sightlines from the DESI spectroscopic dataset are shown by cyan symbols.}
    \label{fig:lofar_image}
\end{figure}

\subsection{Final sample selection}
\label{subsec:sample}

We select radio sources such that their radio end-to-end size ($D_{\rm radio}$) and luminosity ($L_{144}$) properties  are consistent with traditional FRI/FRII systems with double-sided lobes \citep{Hardcastle2020, roy21c}, and satisfying:
\begin{equation}
    4.3 \leq \log_{10}\left(\frac{D_{\rm radio}}{{\rm pc}}\right) \leq 6.5,
\end{equation}
\begin{equation}
    23 \leq \log_{10}\left(\frac{L_{144}}{{\rm W~Hz^{-1}}}\right) \leq 29.5,
\end{equation}
along with positive deconvolved major and minor axes. This yields an initial foreground sample of $90{,}228$ LOFAR-detected radio AGNs with measured radio jet-axis information. 
We then require the foreground radio-galaxy host to have a spectroscopic redshift before crossmatching to the DESI quasar catalog.

For each radio galaxy, we identify the closest background DESI quasar and compute their projected separation ($R_\perp$, or impact parameter) using the radio-galaxy redshift $z_{\rm RG}$. We retain only clean foreground--background configurations satisfying
\begin{equation}
    z_{\rm QSO} > z_{\rm RG} + 0.08,
\end{equation}
and
\begin{equation}
    z_{\rm QSO}-z_{\rm RG} < 2.5.
\end{equation}
The lower limit removes physically associated systems, while the upper limit avoids unrelated intervening spectral features. To exclude host inner-ISM contamination while remaining within the projected radio jet size, we additionally require:
\begin{equation}
    R_\perp > R_{50},
\end{equation}
\begin{align}
R_\perp &< 0.8 D_{\rm{radio}} \nonumber \\
  &< 1.6 R_{\rm{radio}}
\end{align}

where $R_{50}$ is the optical half-light radius of the host, $D_{\rm radio}$ is the projected end-to-end radio size, and $R_{\rm radio}$ is the corresponding radio half-size. The final catalog contains $324$ radio galaxy--quasar pairs spanning $0.05<z_{\rm RG}<1.64$, $0.35<z_{\rm QSO}<3.14$, $R_\perp \simeq 20$--$800$ kpc, and $\log_{10}(L_{144}/{\rm W~Hz^{-1}})\simeq 23.06$--$28.17$.

We compute the position angle from the radio galaxy to the quasar ($\mathrm{PA}_{\rm RG\rightarrow QSO}$) and compare it to the bidirectional LOFAR radio jet axis ($\mathrm{PA}_{\rm radio}$). The angular offset $\theta$ ($0^\circ\leq\theta\leq90^\circ$) is defined as: 
\begin{equation}
    \theta =
    \left| \left[ \left(\mathrm{PA}_{\rm RG\rightarrow QSO}
    - \mathrm{PA}_{\rm radio} + 90^\circ\right)
    \bmod 180^\circ \right] - 90^\circ \right|,
\end{equation}
where small $\theta$ denotes sightlines near the radio axis and large $\theta$ denotes near-transverse sightlines. The main jet-axis stack is defined by $\theta<20^\circ$ and contains $74$ sightlines. 
We also use the normalized impact parameter $R_\perp/D_{\rm radio}$, or $R_\perp/R_{\rm radio}$ to test for radial dependence of our result along the jet axis. 

For each DESI quasar target ID in the final catalog, we retrieve the spectrum using \texttt{SPARCL}. 
Since DESI has finite wavelength coverage, every input foreground--background pair does not contribute to every emission-line stack. Rather than imposing a separate catalog-level redshift cut for each line, the stacking algorithm rejects spectra that do not have finite, positive-inverse-variance pixels in the relevant rest-frame wavelength window. Thus the effective number of spectra contributing to a given stack is set by the wavelength coverage and quality masks at that line.

\subsection{Stacking and uncertainty estimates}
\label{subsec:stacking}

We describe the full details of the continuum subtraction, masking, and noise tests in Appendix~\ref{subsec:stacking}. Briefly, each DESI quasar spectrum is shifted into the foreground radio-galaxy rest frame, we extract a $\pm150$~\AA\ window around H$\alpha$ emission line (rest wavelength $\lambda_0=6562.8$ \AA). We mask strong background-quasar emission lines after projecting them into the foreground radio galaxy rest frame, and subtract a local second-order polynomial continuum fitted outside $|\lambda-\lambda_{\rm H\alpha}|<50$~\AA. The continuum-subtracted residual spectra are then interpolated onto a common rest-frame wavelength grid. Our fiducial stack is an inverse-variance-weighted mean, where we require a minimum of five contributing spectra per wavelength pixel. We measure the integrated H$\alpha$ flux over $|\Delta v|<500~{\rm km~s^{-1}}$ after subtracting a noise baseline (see Appendix~\ref{subsec:noise_control} for details) and thus avoid contamination from [N\,{\sc ii}]. Statistical uncertainties are estimated from 1$\sigma$ scatter of the population distribution, and also from blank-region stacks randomly centered at 6800, 5500, and 4000~\AA. We refer the reader to Appendix ~\ref{subsec:stacking} and \ref{subsec:noise_control} for further details.

\section{Results} \label{sec:results}

We measure stacked \Ha\ fluxes for our sample of quasar-radio galaxy pair, split in bins of projected angular offset from the radio axis ($\theta$), the normalized position along the radio structure ($R_\perp/D_{\rm radio}$, or $R_\perp/R_{\rm radio}$), and stellar mass ($M_\star$). All stacks and figures show inverse-variance-weighted mean fluxes with their corresponding uncertainties unless otherwise mentioned. 

\begin{figure*}[t]
    \centering
    \includegraphics[width=\textwidth]{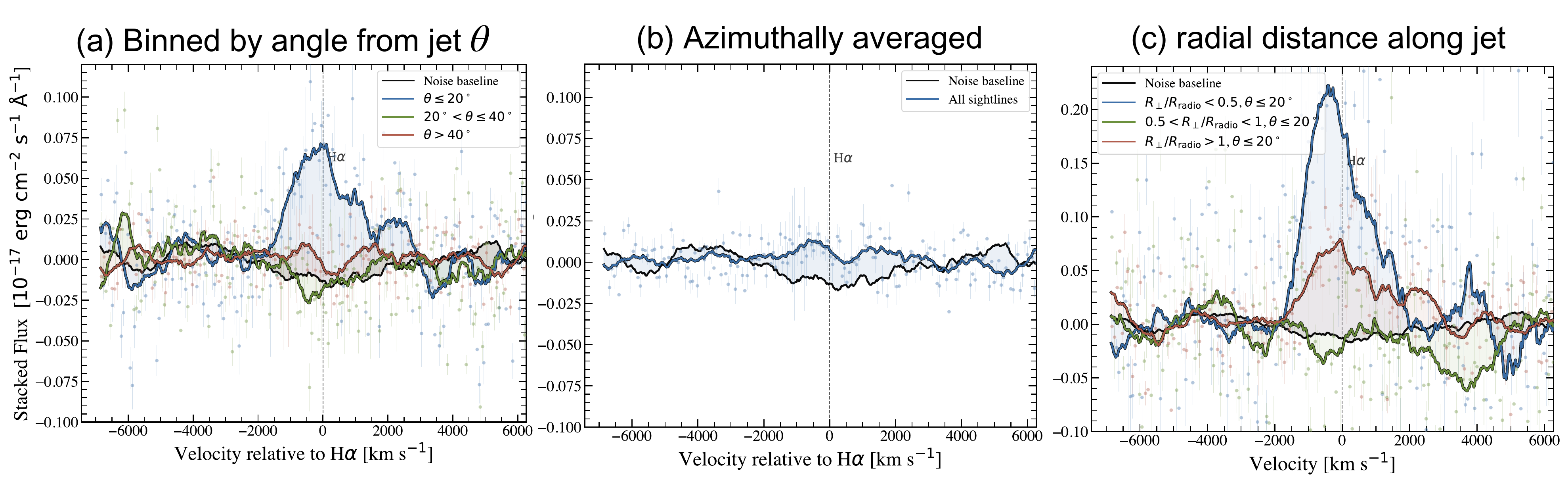}
    \caption{\textit{A:} Angle-binned stacked residual spectra. Sightlines within $\pm\thetajet$ of the radio axis show an \ha excess at the foreground systemic wavelength, while intermediate- and large-angle sightlines do not. The result indicates that the cool ionized \cgm signal is strongly anisotropic and aligned with the projected radio structure.
    \textit{B:} Azimuthally averaged stacked residual \desi spectrum around foreground radio galaxies. The spectrum is shifted to the foreground-galaxy rest frame and centered on \ha. The all-angle stack shows no significant \ha excess, with an integrated significance of $\sigall\sigma$ relative to randomized blank stacks. This null result rules out a simple spherically enhanced \ha halo but does not test for directional jet--\cgm coupling. \textit{C:} Integrated \ha signal as a function of normalized projected distance from the host galaxy for sightlines within $\pm\thetajet$ of the radio axis. The line excess is strongest near the galaxy and again near the projected radio-lobe/cocoon region, suggesting two preferred zones where radio jets couple to cool ionized halo gas.}
    \label{fig:angle_stack}
\end{figure*}

\subsection{A jet-aligned H$\alpha$ excess}\label{subsec:theta}

The left panel of Figure~\ref{fig:angle_stack} (panel a) shows the stacked \Ha \ spectra from the background quasar sightlines, split into three subsamples by their angular offset from the radio jet axis: $\theta < 20^\circ$, $20^\circ<\theta < 40^\circ$ and $\theta > 40^\circ$. We find that the result varies considerably for these different cases. Sightlines within $\pm\thetajet$ of the radio axis, shown in blue, exhibit a strong \ha excess centered near the systemic velocity of the foreground radio galaxies, with a mean integrated flux of  $1.19 \times  10^{-17} \rm \ erg \ cm^{-2} \ s^{-1}$. 
This excess \Ha\ emission is detected with a significance of $> \sigjet\sigma$ above the baseline average in the inverse-variance weighted stacked spectra. 
In contrast, the intermediate and large-angle stacks show no comparable \Ha\ feature.

Such angular dependence of enhanced ionized gas signature  suggests that the detected emission is highly anisotropic and is linked to the radio AGN geometry. This is consistent with current theories that radio jets drive shocks, inflate overpressurized cocoons, and entrain photo-ionized gas preferentially along the jet propagation direction or along cocoon boundaries \citep{begelman89,wagner11, mukherjee18}. In this scenario, the detected \Ha\ emission in the stacked spectra traces dense, localized pockets of cool ionized CGM gas whose emissivity is directionally enhanced along the projected jet axis. However, the same parent sample gives a null detection when the spectra are averaged over all azimuthal angles. Figure~\ref{fig:angle_stack} (middle panel b) shows the stacked spectrum for all background quasar sightlines, independent of their angle relative to the radio source. The azimuthally averaged stack does not show any statistically significant \ha excess.  The integrated signal within the fiducial velocity window is  $\sigall\sigma$ relative to the randomized noise stacks.  

This non detection is consistent with recent absorption-line stacking studies that did not find any strong isotropic enhancement of the cool \cgm  around radio sources relative to matched control samples \citep[e.g.,][]{Chang2024}.
 If the gas responds preferentially along the radio axis, combining all angles dilute the signal, but it produces an excess emission signature when restricted to sightlines aligned with the projected radio source, as shown above.

We perform several checks to test whether the angular signal could be produced by a small number of outliers or by residual systematics in the quasar spectra. We find that the integrated flux changes only by $\sim 4-7$\% after removing individual high-weight spectra one a a time using the jack-knife method of rejection, thus retaining the detection significance to be $>4.7 \sigma$. The signal also cannot be reproduced artificially by stacking randomized foreground redshifts, or from any combination of nearby blank spectral regions. Our tests indicate that the detected \Ha\ flux is physical and emerges from combining true foreground redshift and the observed radio-axis geometry. 
A qualitatively similar angular dependence is seen in the SIMBA simulation from a representative jet-mode galaxy shown in Appendix~\ref{app:sims}, where the area-averaged H$\alpha$ surface brightness is largest close to the projected jet axis and decreases toward larger azimuthal angles.




\subsection{Radial variation of \Ha \  along the radio-jet axis}

Next, we investigate how the \ha \ emission fluxes change as a function of projected radii from the center along the radio lobes. 
Figure~\ref{fig:angle_stack} (right, panel c) shows the inverse-variance weighted stacked spectra measured in bins of projected radial distance of the sightline scaled by radio half-size, $R_\perp/R_{\rm radio}$, for the subsample aligned close to the jet ($|\theta|<\thetajet$). So keeping angular offset from the radio jet to be fixed and within $\thetajet$, we split our sample to $R_\perp/R_{\rm radio} < 0.5$, $ 0.5 < R_\perp/R_{\rm radio} < 1.0$, and $R_\perp/R_{\rm radio} > 1.0$.
The goal is to quantify how the flux changes as the sightline moves out from the center towards the lobe terminus. We find that stacked \ha is not enhanced uniformly along the radio jet on average. The emission is strongest and is detected with $> 7 \sigma$ near the central regions close to the host galaxy where the integrated \Ha \ flux = 3.46 $\times  10^{-17} \rm \ erg \ cm^{-2} \ s^{-1}$. The mean \Ha\ signal again start to rise toward larger radii near the radio lobes with a mean flux of 1.31 $\times  10^{-17} \rm \ erg \ cm^{-2} \ s^{-1}$. Projected distances at intermediate radii show weaker emission or no detection. Thus, the \Ha\ emission appears to be concentrated in those regions where jet-gas coupling is expected to be most effective. Near the host galaxy (but beyond $R_{50}$ to avoid the ISM emission, see \S \ref{subsec:sample}), the \ha emission likely traces gas at the transition between the dense ISM and the inner CGM, where the young jet first encounters the surrounding halo. The increased flux at higher radii where $R_\perp/R_{\rm radio} > 1.0$ is possibly associated with the lobe or cocoon interface, where the jet deposits mechanical energy into the ambient CGM and can compress or ionize cool gas.

This interpretation is consistent with resolved studies of powerful radio galaxies at higher redshift ($z>3.5$). Spatially resolved IFU observations from JWST/NIRSpec have found extended nebular emission in \Ha, \oiii, \oii, and other rest-frame optical lines that are strongly aligned with their radio jet axes  \citep[e.g.,][]{roy24,roy25, wang24, saxena24}. Thus the ionized gas emission exhibit distinct enhancements along the compact radio lobes from a few to several tens of kpcs in high-redshift systems. The systems studied in this work extend this experiment to lower-redshift radio galaxies with larger and more evolved radio structures, where lobe-to-lobe radio sizes $D_{\rm radio}$ reach $\gtrsim 300$ kpc. Our stacked measurements therefore suggest that the same energy coupling scenario seen in relatively compact high-redshift systems may persist to CGM scales in nearby radio galaxies, near  the sites where radio lobes interact with the surrounding medium.



\section{Discussions}
\subsection{Physical Constraints from the Stacked H$\alpha$ Signal}
\label{subsec:constraints}

The detected \ha flux provides a direct constraint on the amount of cool ionized gas produced by jet--\cgm coupling. For a characteristic background quasar sightline passing through the foreground radio galaxy CGM at luminosity distance $D_L$, the integrated \ha \ line luminosity is
\begin{equation}
    L_{\rm H\alpha} = 4\pi D_L^2 F_{\rm H\alpha},
\end{equation}
where $F_{\rm H\alpha}$ is the measured integrated flux. If we consider only the weighted mean flux from the sightlines lying along the radio jet axis $\theta < \thetajet$ (irrespective of their radial distance) from \S\ref{subsec:theta}, $F_{\rm H\alpha}$ = \fha $\times  10^{-17} \rm \ erg \ cm^{-2} \ s^{-1}$. 
Under Case B recombination, the ionized gas mass associated with the emitting phase can be written as

\begin{align}
M_{\rm ion} &\simeq
    \frac{m_p L_{\rm H\alpha}}
    {n_e \alpha_{\rm H\alpha}^{\rm eff} h\nu_{\rm H\alpha}} \\
    &\approx3.2\times10^9
\left(\frac{L_{\rm H\alpha}}{10^{40}\ {\rm erg\ s^{-1}}}\right)
\left(\frac{10^{-2}\ {\rm cm^{-3}}}{n_e}\right)
M_\odot,  
\end{align}

where $n_e$ is the electron density normalized by $10^{-2}\rm \ cm^{-3}$, $m_p$ is the mass of a proton and $\alpha_{\rm H\alpha}^{\rm eff} \rm = \   1.17 \times 10^{-13} \rm \ cm^3 \ s^{-1}$ is the effective Case B recombination coefficient at $T\simeq10^4$ K. 
The normalization $n_e=10^{-2}\ {\rm cm^{-3}}$ is motivated by approximate pressure balance between the cool \ha-emitting phase and a surrounding hot halo. For a hot CGM phase with $T_{\rm hot}\sim10^6$ K and $n_{\rm hot}\sim10^{-4}\ {\rm cm^{-3}}$ \citep{Fumagalli2024}, the thermal pressure is
\begin{equation}
    \frac{P}{k_{\rm B}} \simeq n_{\rm hot}T_{\rm hot}
    \sim 10^2\ {\rm K\ cm^{-3}} .
\end{equation}
If the $T_{\rm cool}\sim10^4$ K ionized hydrogen clouds are in rough pressure equilibrium with this hotter ambient medium, then an order-of-magnitude cool-phase CGM density scale can be estimated as:
\begin{equation}
    n_{\rm cool}
    \simeq
    n_{\rm hot}\frac{T_{\rm hot}}{T_{\rm cool}}
    \sim
    10^{-2}\ {\rm cm^{-3}} .
\end{equation}

Thus, assuming $n_e=10^{-2}\ {\rm cm^{-3}}$, and using the mean integrated flux $F_{\rm H\alpha}$ measured for the jet-aligned sightlines at $\langle z_{\rm RG}\rangle=0.66$, we obtain a characteristic luminosity $L_{\rm H\alpha}=2.2\times10^{40}\ {\rm erg\ s^{-1}}$. The corresponding ionized gas mass is $7\times 10^9 \ \rm M_{\odot}$.
Thus, even a faint stacked \ha signal can correspond to a substantial cool ionized gas reservoir at plausible CGM densities.

The emission line luminosity can be powered by several mechanisms including AGN photoionization, shock ionization along the jet axes, turbulent mixing layers, or collisional excitation in cooling gas. Our stacked measurements do not distinguish between these channels. However, the alignment with the radio axis and the radial concentration near the nucleus and outer radio structures point to a clear mechanical connection with the jet.


\subsection{Comparison with H$\alpha$ CGM constraints in star-forming galaxies}
\label{subsec:comparison_sf}

\begin{figure}[t]
    \centering
    \includegraphics[width=0.49\textwidth]{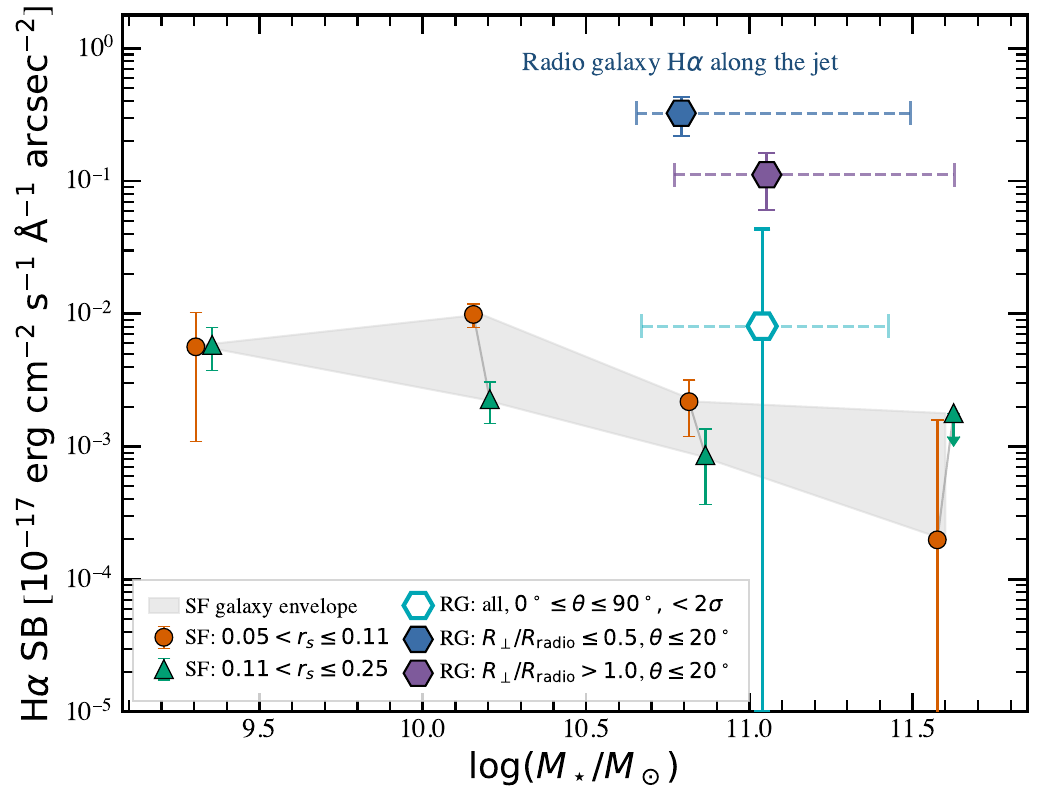}
    \caption{H$\alpha$ emission surface-brightness as a function of stellar mass. All measurements are expressed as rest-frame surface brightnesses after correcting for cosmological dimming by a factor of $(1+z)^4$. The orange circles and green triangles show star forming galaxy measurements from \citet{Zhang2020} in two bins of scaled projected radius, $r_s=r_p/R_{180}$, where $r_p$ is the projected distance from the galaxy center and $R_{180}$ is the virial radius. The gray shaded region marks the corresponding range of H$\alpha$ surface brightnesses observed from the inner to outer regions of these star forming galaxies and groups \citep{Zhang2016, Zhang2020}. The cyan hexagon shows the radio-galaxy stack including all background-quasar sightlines, independent of angular position relative to the radio axis ($0^\circ\leq\theta\leq90^\circ$). 
   Since the \ha\ flux stacked over all azimuthal angles is a non-detection ($<2\sigma$; see Figure~\ref{fig:angle_stack} panel b), this point should be interpreted as an upper limit. This upper limit is consistent with the H$\alpha$ surface brightness expected for normal galaxy halos \citep{Zhang2016, Chang2024}. The blue and purple hexagons show radio-galaxy stacks restricted to sightlines within $20^\circ$ of the projected radio axis. They are separated into inner sightlines with $R_\perp/R_{\rm radio}\leq0.5$ and outer sightlines with $R_\perp/R_{\rm radio}>1.0$ which are close to the radio lobes.  The \Ha\ emission from jet-aligned sightlines are much stronger, and are enhanced by nearly one to two orders of magnitude relative to the normal galaxy baseline at similar stellar mass. This indicates that the \Ha\ emissivity from cool ionized CGM is preferentially brightened along directions where radio jets interact with halo gas.}
    \label{fig:ha_mstar}
\end{figure}


Stacked recombination-line measurements have recently provided a useful baseline for the level of cool ionized gas expected around normal star forming galaxies. Using SDSS spectra, \citet{Zhang2016} detected stacked \ha+\NII\ emission around low-redshift galaxies from projected radii of $\sim5$ kpc to beyond $100$ kpc, with a surface-brightness profile declining approximately as $r_p^{-1.9\pm0.4}$. Extending this analysis to a larger sample of low-redshift isolated galaxies, groups, and clusters, \citet{Zhang2020} found that the contribution of the cool $T\sim10^4$ K CGM fraction decreases strongly with stellar and halo mass. Their measurements provide an empirical baseline for the faint, approximately halo-averaged \ha\ emission expected around normal massive, evolved galaxy systems.

Figure~\ref{fig:ha_mstar} measures the H$\alpha$ surface brightnesses corrected by cosmological dimming factor of $(1+z)^4$  as a function of stellar mass, and compares the  measurements for non-AGN galaxy halos from \citet{Zhang2020} with our radio-galaxy measurements. The comparison samples are not exactly identical because the Zhang et al. measurements are stacked in bins of scaled projected radius, $r_s=r_p/R_{180}$, where $r_p$ is the projected distance from the galaxy and $R_{180}$ is the virial radius. On the other hand, our radio galaxy sample are stacked in bins of normalized radial distance from the center relative to the radio jet length, ${\rm R_\perp}/R_{\rm radio}$, where ${\rm R_\perp}$ is the projected distance of the sightline equivalent to the impact parameter and $R_{\rm radio}$ is the size of one side of the radio jet. The stacks in our sample shown here are further split by the angular offset of the sightline from the projected jet axis ($\theta$). However, both measurements are expressed as stacked H$\alpha$ surface brightness and provide a direct comparison between the cool ionized CGM clouds prevalent in these systems.

 Earlier, we found that when all radio-galaxy sightlines are combined over all azimuthal angles with respect to the jet ($0^\circ\leq\theta\leq90^\circ$), the stacked spectrum yields no significant \ha\ detection, with an integrated significance of $<2\sigma$ as shown in Figure \ref{fig:angle_stack} panel b. Therefore, the corresponding cosmological-dimming-corrected surface-brightness value shown by the cyan symbol in Figure~\ref{fig:ha_mstar} should be interpreted as an upper limit rather than a detection. 
This upper limit is consistent with the H$\alpha$ surface brightness expected for normal galaxy halos \citep{Chang2024, Zhang2016} and indicate that their globally averaged cool ionized CGM is possibly no different than that of actively star-forming halos. 
But the \Ha\ emission becomes drastically stronger when the stacks are restricted to sightlines aligned with the radio jet axis. Within $\theta\leq20^\circ$ around the radio jets, both the inner ($R_\perp/R_{\rm radio}\leq0.5$) and outer ($R_\perp/R_{\rm radio}>1.0$) radial bins (blue and purple hexagons respectively) show substantially enhanced H$\alpha$ surface brightnesses  compared to the \citet{Zhang2020} measurements at similar stellar mass. The jet-aligned measurements are enhanced by 1--2 orders of magnitude relative to the normal-galaxy baseline at similar stellar mass. Hence the impact of jets on the CGM are about two orders of magnitude stronger than similar impact driven by star forming galaxies.

The non-detections when averaged over all azimuthal angle do not establish a globally larger H$\alpha$-emitting reservoir in low redshift radio-galaxy halos. Instead, their detected enhanced \ha emission is localized along the radio jet axis, that shows that the gas is not distributed or excited isotropically. The jet modifies the emissivity of cool CGM gas in a strongly directional manner. This directional enhancement can arise through two related physical pathways. The radio jet may encounter pre-existing cool CGM clouds and increase their H$\alpha$ emissivity through compression, shock ionization, or turbulent mixing at the lobe-halo interface. Alternatively, some of the \ha \ emitting material may originate in the ISM or inner CGM and be entrained outward by the outflow. These two scenarios may not be mutually exclusive from each other. Since recombination-line emissivity scales approximately as $n_e^2$, locally compressed or more highly ionized clouds can become dramatically brighter in \ha\ without requiring an increase in the total cool-gas mass averaged over the halo. The comparison with normal galaxies clarifies the role of radio-mode feedback in these systems.  The enhanced \ha\ emission close to the radio axis, detected both near the host and toward the outer lobes, provides direct evidence that radio jets affect the surrounding ISM and CGM in a strongly anisotropic manner.

\subsection{Energetics estimates}
The rest-frame 1.4~GHz radio luminosities in our sample ranges between $\log L_{\rm 1.4\,GHz} \simeq39$--$43 \ {\rm erg~s^{-1}}$, corresponding to inferred jet mechanical powers of $\log P_{\rm jet}\simeq42.2$--$45.5  \ {\rm erg~s^{-1}}$.
For a mean $\langle \log P_{\rm jet}\rangle\simeq43.5 \ {\rm erg~s^{-1}}$ \citep{roy25}, the available mechanical power is at least an order of magnitude larger than the characteristic mechanical energy injection from stellar feedback in normal star-forming galaxies. For supernovae and stellar winds, the characteristic energy deposition rate is recorded to be $\dot E_{\rm SF}\sim7\times10^{41}({\rm SFR}/1\,M_\odot\,{\rm yr^{-1}})\ {\rm erg~s^{-1}}$  \citep[e.g.,][]{Leitherer1999,Veilleux2005}. Thus, the factor of 1--2 orders of magnitude enhancement in jet-aligned \ha\ surface brightness relative to the star-forming-galaxy halo baseline is energetically feasible.


Using the observed jet-aligned luminosity, $L_{\rm H\alpha}\simeq2\times10^{40}\ {\rm erg~s^{-1}}$, we find
\begin{equation} \nonumber
    \frac{L_{\rm H\alpha}}{\langle P_{\rm jet}\rangle}
    \simeq
    6\times10^{-4}
    \simeq
    0.06\%.
\end{equation}
Thus, only of order $10^{-3}$ of the available jet mechanical power needs to emit as \ha\ radiation from the cool ionized phase to account for the observed enhancement in \ha emission. This ratio is only a small fraction of the total coupling efficiency of the jet to the CGM, as \ha\ traces gas only at a temperature of $T\sim10^4$~K gas. This low required energy coupling efficiency is also consistent with a representative jet-mode AGN from the SIMBA cosmological simulation demonstrated in Appendix~\ref{app:sims}. 
Additional energy from the jet may be deposited as kinetic or thermal energy in other temperature and gas phases, most predominantly in the shock-heated X-ray-emitting gas with $T\gtrsim 10^6$K \citep{mcnamara07, mcnamara12, fabian12, Gitti2012}.

\subsection{The origin of the \ha ionized gas: insights from \mgii absorption profiles}
\label{subsec:fate_coolgas}

\begin{figure*}[t]
    \centering
    \includegraphics[width=\textwidth]{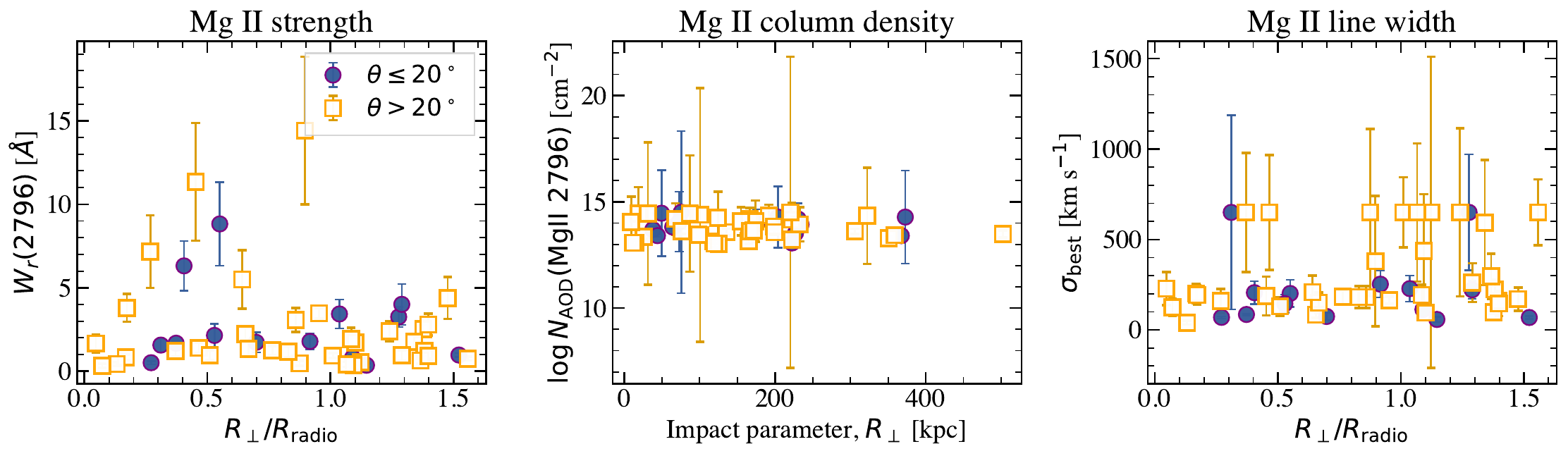}
    \caption{Mg\,{\sc ii} absorption properties for individual radio-galaxy sightlines. 
Blue filled circles show absorbers along jet-aligned sightlines with 
$\theta\leq20^\circ$, while orange open squares show absorbers at 
$\theta>20^\circ$. \textit{Left:} Mg\,{\sc ii} $\lambda2796$ rest-frame equivalent 
width as a function of normalized projected distance, $R_\perp/R_{\rm radio}$. 
\textit{Center} apparent-optical-depth Mg\,{\sc ii} $\lambda2796$ column density as a 
function of projected impact parameter, $R_\perp$. \textit{Right:} fitted velocity width 
as a function of $R_\perp/R_{\rm radio}$. The jet-aligned and off-axis absorbers 
span similar ranges in equivalent width, column density, and line width, 
indicating that the Mg\,{\sc ii}-bearing cool clouds are not strongly enhanced 
or kinematically distinct along the radio-jet direction. This contrasts with 
the strongly jet-aligned H$\alpha$ emission, suggesting that radio jets primarily 
brighten or ionize a subset of a more broadly distributed cool CGM reservoir.
}
    \label{fig:mgii}
\end{figure*}

\begin{figure}[t]
    \centering
    \includegraphics[width=0.49\textwidth]{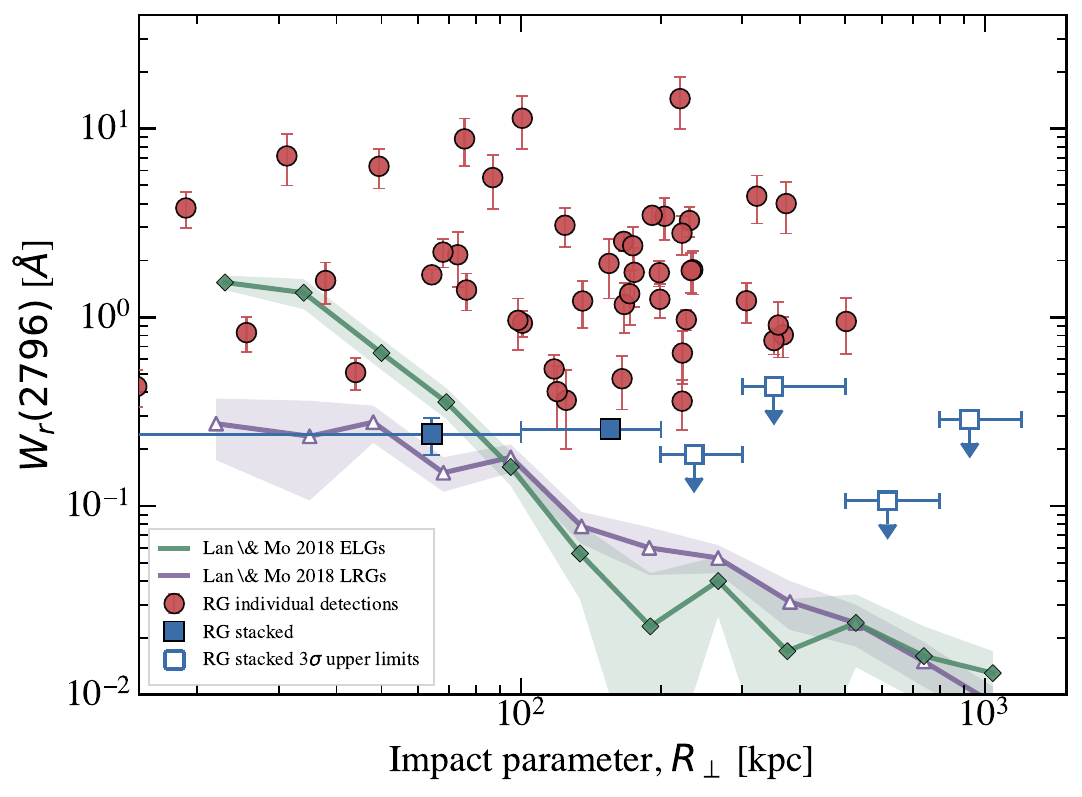}
    \caption{
Mg\,{\sc ii} $\lambda2796$ rest-frame equivalent width as a function of projected impact parameter, $R_\perp$. Red circles show individual Mg\,{\sc ii} detections associated with the radio-galaxy sightlines in this work. Filled blue squares show stacked measurements, while open blue squares indicate $3\sigma$ upper limits from stacks without individual detections. For comparison, the green and purple curves show the average Mg\,{\sc ii} absorption profiles around emission-line galaxies (ELGs) and luminous red galaxies (LRGs) from \citet{Lan2018}, respectively, with shaded regions indicating the reported uncertainties. Individual radio-galaxy absorbers are detected over a broad range of projected radii, including beyond $\sim100$ kpc, and several lie above the average LRG profile. However, the stacked signal is weak and several radial bins yield only upper limits, indicating that the cool Mg\,{\sc ii}-bearing gas around radio galaxies is extended and clumpy, with a low covering fraction rather than a smooth, high-covering-fraction halo component.
}
    \label{fig:mgii_lm18}
\end{figure}

The previous sections show that the azimuthally averaged H$\alpha$ stack is not significantly detected, while the same sample shows a strong H$\alpha$ excess when restricted to sightlines close to the projected radio axis.
 This raises a key question: does the jet-aligned \ha\ excess reflect an intrinsically larger amount of cool gas along the radio jet axis, or does the jet locally brighten a more broadly distributed cool-gas reservoir? 

To address this, we search for Mg\,{\sc ii} $\lambda\lambda2796,2803$ absorption in the same DESI background-quasar sightlines. Mg\,{\sc ii} provides a complementary probe because it is sensitive to the presence and column density of cool, metal-enriched clouds, rather than to the $n_e^2$ emissivity that governs recombination emission. We find candidate Mg\,{\sc ii} absorbers in $\simeq22\%$ of the jet-aligned sightlines ($\theta<20^\circ$), compared with $\simeq16\%$ of the off-axis sightlines ($20^\circ<\theta<90^\circ$). The difference in detection is moderate. Moreover, Figure~\ref{fig:mgii} demonstrates that the Mg\,{\sc ii} equivalent widths (left panel), apparent-optical-depth column densities (middle panel) and fitted line-widths (right panel) are broadly similar for jet-aligned sightlines ($\theta\leq20^\circ$; blue circles) and off-axis sightlines ($\theta>20^\circ$; yellow squares). These quantities also show no clear systematic trend with normalized distance along the radio source for our sample. Thus, unlike \ha, the Mg\,{\sc ii} clouds appear to be distributed over a broad range of azimuthal angle
and does not show any strong evidence for being preferentially concentrated along the projected jet axis.

This result is consistent with previous absorption-line work showing that massive quiescent galaxies can retain substantial cool CGM reservoirs. For example, COS-Halos survey detected cool, largely bound \ion{H}{1}-bearing gas around passive early-type galaxies, with inferred cool CGM masses of $10^9$--$10^{11}\,M_\odot$ \citep{Thom2012}. Mg\,{\sc ii} absorption has also been detected statistically around luminous red galaxies and other massive halos \citep{Zhu2014,Lan2018}. In our sample, Figure~\ref{fig:mgii_lm18} shows that individual Mg\,{\sc ii} absorbers, shown in red symbols, are found to large projected distances, often beyond $\sim100$ kpc and in some cases above the average luminous-red-galaxy profiles from \citet{Lan2018}. However, the stacked Mg\,{\sc ii} signals shown in blue squares are weak, with several radial bins yielding only upper limits. The combined evidence points to an extended, clumpy cool phase with low covering fraction, rather than a smooth, high-covering-fraction reservoir.

This contrast between the nature of \Ha \ emission and Mg\,{\sc ii} absorption can occur if radio jets propagate through an extended population of isotropically distributed cool clouds and locally alter their physical state. 
Jet-driven cocoons can create overpressured regions relative to the ambient medium \citep{begelman89}, and simulations of jets propagating through clumpy multiphase gas show that the expanding jet-driven cocoon and backflow can compress cool clouds, drive shocks and turbulence, and increase the temperature and ionization state of the gas along the jet path \citep{Sutherland2007,wagner12, mukherjee16, mukherjee18,dutta24}.
Since \ha\ emissivity scales approximately as $j_{\rm H\alpha}\propto n_e^2$, even a modest density enhancement in a small subset of clouds can produce a large increase in recombination emission without a comparable increase in the Mg\,{\sc ii} covering fraction. This causes a large enhancement of \ha \ signal in the jet direction, but no dependence of the Mg\,{\sc ii} signal with azimuthal angle. Another possibility is gas being entrained along the jet as shown by  powerful high-redshift radio galaxies in tens of kpc scale \citep[e.g.,][]{roy24, roy25}. We indeed observe preliminary signatures of a significantly broadened, stacked \Ha\ profile that spans $\sim2000$ km/s that may suggest kinematic disturbance or entrainment, but this needs a more detailed followup study. 

This favors a picture in which Mg\,{\sc ii}-bearing clouds provide an extended, clumpy cool-gas reservoir from which at least part of the H$\alpha$ emission arises. The radio jet then increases the density, pressure, and ionization state of the subset of clouds along its path, making them disproportionately bright in H$\alpha$. Some entrainment from the ISM or inner CGM may also occur, but the lack of an enhanced Mg\,{\sc ii} velocity width or velocity offset for jet-aligned sightlines suggests that the dominant effect is local brightening or ionization of pre-existing CGM clouds.

\section{Conclusions} \label{sec:conclusion}

We have presented a statistical search for faint H$\alpha$ emission from the cool ionized CGM around radio galaxies by stacking DESI background-quasar spectra at the redshifts of foreground LoTSS radio galaxies. The key feature of our experiment is that we preserve the projected geometry of each sightline relative to the radio axis, allowing us to test whether radio jets affect the cool CGM anisotropically rather than only in an azimuthally averaged sense. Our main results are:

\begin{enumerate}
    \item The H$\alpha$ signal is strongly dependent on projected radio-jet alignment. Sightlines within $\theta<20^\circ$ of the radio axis show a clear H$\alpha$ excess centered near the systemic velocity of the foreground radio galaxies, with a mean integrated flux of $F_{\rm H\alpha}=1.19\times10^{-17}\ {\rm erg\ cm^{-2}\ s^{-1}}$ and a detection significance of $>5\sigma$. In contrast, intermediate and large-angle sightlines show no comparable emission feature.
    The azimuthally averaged radio-galaxy stack also does not show a significant H$\alpha$ detection. When all sightlines are combined over $0^\circ\leq\theta\leq90^\circ$, the integrated signal is $<2\sigma$ relative to the noise/control stacks. Thus, the \Ha \ emitting clouds are clumpy and are distributed anisotropically, and lie preferentially enhanced along the projected radio axis.

    \item The jet-aligned H$\alpha$ emission is also radially structured. For sightlines within $\theta<20^\circ$, the strongest emission occurs close to the host galaxy at the ISM-CGM boundary, beyond the optical half-light radius, with $F_{\rm H\alpha}=3.46\times10^{-17}\ {\rm erg\ cm^{-2}\ s^{-1}}$. The signal weakens at intermediate normalized radius and rises again near the outer radio-lobe region, with $F_{\rm H\alpha}=1.31\times10^{-17}\ {\rm erg\ cm^{-2}\ s^{-1}}$. This suggests two preferred zones of jet--gas coupling: near the ISM--CGM transition and near the radio lobe--CGM interface.

    \item Compared with stacked H$\alpha$ measurements around normal star-forming galaxies and galaxy systems from \citet{Zhang2016,Zhang2020}, the radio-galaxy spectra stacked over all angles with respect to the jet lies within the expected massive-halo baseline, while the jet-aligned stacks are elevated by almost two orders of magnitude. This contrast shows that the radio-galaxy halos are not globally richer in \Ha, but radio jets locally enhance the H$\alpha$ emissivity of cool CGM material along preferred directions.

    \item The inferred mass and energetics are consistent with a jet-powered origin for the enhanced H$\alpha$ emission. For the characteristic jet-aligned luminosity, $L_{\rm H\alpha}\simeq2.2\times10^{40}\ {\rm erg\ s^{-1}}$, the implied ionized gas mass is $\sim7\times10^9\,(10^{-2}\ {\rm cm^{-3}}/n_e)\ M_\odot$, where the assumed density is motivated by pressure balance between $T\sim10^4$ K clouds and a $T\sim10^6$ K hot halo. The required H$\alpha$ luminosity is only $\sim6\times10^{-4}$ of the mean inferred jet mechanical power. Thus, a very small fraction of the available jet energy is sufficient to produce the observed H$\alpha$ excess in the cool ionized CGM.

    \item Mg\,{\sc ii} absorption provides a complementary view of the cool CGM. Candidate Mg\,{\sc ii} absorbers are found along $\simeq22\%$ of jet-aligned sightlines and $\simeq16\%$ of off-axis sightlines, with no strong difference in equivalent width, column density, velocity offset, or line width. Individual Mg\,{\sc ii} absorbers can extend beyond $\sim100$ kpc, but the stacked signal is weak, indicating a clumpy, low-covering-fraction cool phase. The contrast between broadly distributed Mg\,{\sc ii} absorption and strongly jet-aligned H$\alpha$ emission suggests that jets do not create all of the cool gas, but instead brighten, ionize, compress, or perturb a subset of pre-existing cool CGM clouds.

\end{enumerate}

These results provide direct statistical evidence that radio jets affect the cool CGM anisotropically. The \ha \ around radio galaxies, when globally averaged over all angles is not significantly detected and remains consistent with the weak halo-averaged CGM emission expected for massive quiescent systems. The radio source instead imprints a directional H$\alpha$ flux excess that traces where jet-driven cocoons and lobes compress, ionize, entrain, or mix cool clouds. The coexistence of broadly distributed Mg\,{\sc ii} absorption and strongly jet-aligned H$\alpha$ emission suggests that radio jets make a subset of an extended, clumpy CGM reservoir radiatively visible. This offers a multiphase view of radio-mode feedback operating beyond the ISM and into the halo gas that regulates future accretion and star formation.
A qualitatively similar behavior is observed in the cosmological large volume simulation SIMBA in a jet-mode AGN host galaxy, as discussed in Appendix~\ref{app:sims}. Future simulation-observation comparisons that forward-model DESI-like sightlines through statistically matched samples of jet-mode galaxies will be essential for turning this qualitative agreement into quantitative constraints on where, when, and how radio jets deposit energy into cool halo gas.

\begin{acknowledgements}

N.R. acknowledges support from the Exploration Postdoctoral Fellowship at the School of Earth and Space Exploration (SESE), Arizona State University.
NR and SB thank Prof. Romeel Davé and Dr. Tianyi Yang for helping with the analyses of the SIMBA simulations. 
This research used data obtained with the Dark Energy Spectroscopic Instrument (DESI). DESI construction and operations is managed by the Lawrence Berkeley National Laboratory. This material is based upon work supported by the U.S. Department of Energy, Office of Science, Office of High-Energy Physics, under Contract No. DE–AC02–05CH11231, and by the National Energy Research Scientific Computing Center, a DOE Office of Science User Facility under the same contract. Additional support for DESI was provided by the U.S. National Science Foundation (NSF), Division of Astronomical Sciences under Contract No. AST-0950945 to the NSF’s National Optical-Infrared Astronomy Research Laboratory; the Science and Technology Facilities Council of the United Kingdom; the Gordon and Betty Moore Foundation; the Heising-Simons Foundation; the French Alternative Energies and Atomic Energy Commission (CEA); the National Council of Humanities, Science and Technology of Mexico (CONAHCYT); the Ministry of Science and Innovation of Spain (MICINN), and by the DESI Member Institutions: www.desi.lbl.gov/collaborating-institutions. The DESI collaboration is honored to be permitted to conduct scientific research on I’oligam Du’ag (Kitt Peak), a mountain with particular significance to the Tohono O’odham Nation. Any opinions, findings, and conclusions or recommendations expressed in this material are those of the author(s) and do not necessarily reflect the views of the U.S. National Science Foundation, the U.S. Department of Energy, or any of the listed funding agencies.
This research also utilized data from the LOFAR Two-metre Sky Survey. LOFAR is the Low Frequency Array designed and constructed by ASTRON. It has observing, data processing, and data storage facilities in several countries, which are owned by various parties (each with their own funding sources), and which are collectively operated by the LOFAR ERIC under a joint scientific policy. The LOFAR resources have benefited from the following recent major funding sources: CNRS-INSU, Observatoire de Paris and Université d'Orléans, France; BMFTR, MKW-NRW, MPG, Germany; Science Foundation Ireland (SFI), Department of Business, Enterprise and Innovation (DBEI), Ireland; NWO, The Netherlands; The Science and Technology Facilities Council, UK; Ministry of Science and Higher Education, Poland; The Istituto Nazionale di Astrofisica (INAF), Italy.

\end{acknowledgements}

\section{Appendix}
\subsection{A Simulated H$\alpha$ View of Jet-mode Feedback from the SIMBA Simulation}
\label{app:sims} 

\begin{figure*}[t]
    \centering
    \includegraphics[width=0.98\textwidth]{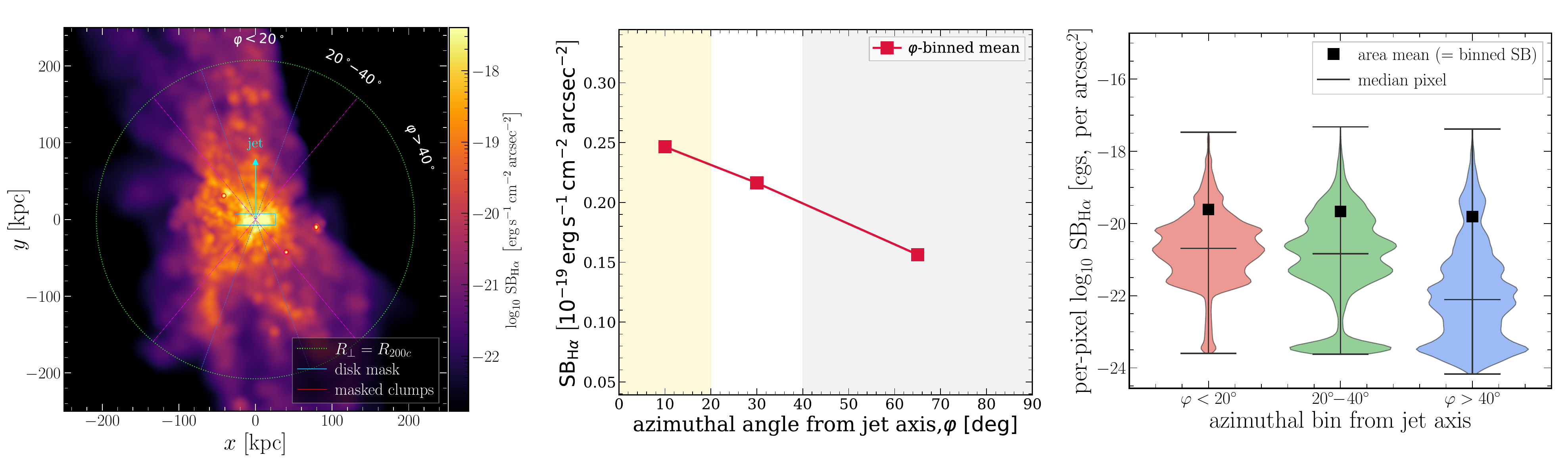}
    \caption{ Intrinsic H$\alpha$ surface-brightness diagnostics of a jet-mode galaxy (\texttt{g54} in \texttt{m25n512}) from the \texttt{SIMBA} simulation, viewed edge-on to the jet. \textit{Left:} H$\alpha$ surface-brightness map over a $\pm250\,\mathrm{kpc}$ field. The cyan arrow marks the projected jet axis, the green dotted circle marks the $R_\perp=R_{200c}$ aperture, the blue rectangle marks the masked edge-on disk, and the red circles mark masked compact star-forming clumps. The blue and magenta radial lines show the three azimuthal bins measured from the jet axis: $\varphi<20^\circ$, $20^\circ<\varphi<40^\circ$, and $\varphi>40^\circ$. \textit{Center:} Stacked H$\alpha$ surface brightness as a function of azimuthal angle $\varphi$. Red squares show the mean surface brightness in the three coarse azimuthal bins, matching the angular bins used for the observed \Ha\ stacks. \textit{Right:} Per-pixel $\log_{10}$ H$\alpha$ surface-brightness distributions in the same three azimuthal bins. Black squares show the area-averaged surface brightness, and horizontal black lines show the median pixel value. The means lie above the medians, indicating that the azimuthally binned signal is dominated by relatively small number of bright pixels.}
    \label{fig:sims}
\end{figure*}

To complement the observed results with a physical picture of the cool circumgalactic gas around a radio-jet host, we construct a mock H$\alpha$ surface-brightness map of a representative galaxy with a jet-mode AGN chosen from the SIMBA cosmological hydrodynamic simulation \citep{Dave2019}. We use SIMBA's high-resolution $25\,h^{-1}\,\mathrm{Mpc}$ volume, with $2\times512^3$ resolution elements, at $z=0$. The galaxy is selected from the jet-active sample of \citet{Tainyi2024}. We follow that work for the jet-mode selection criteria and for the definition of the jet axis. The selected central galaxy has host galaxy properties matched with our sample, with $\log(M_\star/M_\odot)=10.6$, $\log(M_{\rm BH}/M_\odot)=7.6$, $f_{\rm Edd}\simeq5\times10^{-3}$, and $R_{200c}=208\,\mathrm{kpc}$. 
For each gas resolution element, we assign an H$\alpha$ volume emissivity  \begin{equation}      \varepsilon_{{\rm H}\alpha,i}      =      \varepsilon_{{\rm H}\alpha}(n_{{\rm H},i},T_i,Z_i),  \end{equation}  from a CLOUDY v23.01 \citep{Chatzikos2023} photoionization-plus-recombination grid computed under the \citet{Haardt2012} ultraviolet background. We integrate the emissivity along the line of sight with \textsc{yt} \citep{Turk2011} to obtain the intrinsic H$\alpha$ intensity,  \begin{equation}      I_{{\rm H}\alpha}(\mathbf{X})      =      \int \varepsilon_{{\rm H}\alpha}\,d\ell,  \end{equation}  where $\mathbf{X}$ is the projected sky position. Assuming isotropic emission, the intrinsic, rest-frame surface brightness is  \begin{equation}      \mathrm{SB}_{{\rm H}\alpha}(\mathbf{X})      =      \frac{I_{{\rm H}\alpha}(\mathbf{X})}{4\pi}      \left(\frac{1\,\mathrm{sr}}{4.255\times10^{10}\,\mathrm{arcsec}^{2}}\right),  \end{equation}  in units of $\mathrm{erg\,s^{-1}\,cm^{-2}\,arcsec^{-2}}$. The galaxy is viewed edge-on to the jet, with an integration depth of $\pm R_{200c}$. To isolate the diffuse CGM, we measure the area-averaged (stacked) H$\alpha$ surface brightness within $R_\perp<R_{200c}$ after masking the disk emission and removing the contribution from three compact star-forming clumps which may not be associated with the main galaxy. 
The remaining emission is measured in three azimuthal bins from the projected jet axis: $\varphi<20^\circ$, $20^\circ<\varphi<40^\circ$, and $\varphi>40^\circ$, matching the angular bins used for the observed stacks. 

In this representative galaxy with an active jet-mode AGN, the diffuse H$\alpha$ surface brightness is clearly anisotropic. Figure~\ref{fig:sims} shows that the brightest H$\alpha$-emitting structures are preferentially located near the projected jet axis. The area-averaged surface brightness exhibit the largest mean surface brightness value close to the jet axis in the $\varphi<20^\circ$ angular bin, and declines with increasing angle from the jet towards the disk plane with $\mathrm{SB}\simeq2.5$, $2.2$, and $1.6\times10^{-20}\,\mathrm{erg\,s^{-1}\,cm^{-2}\,arcsec^{-2}}$ in the three azimuthal bins, respectively (Figure \ref{fig:sims}, middle panel). 
The trend is qualitatively consistent with our observed stacks of \Ha \ from the DESI spectra, where the detected H$\alpha$ signal is strongest for sightlines closest to the projected radio axis. Interestingly, the  surface-brightness distribution per-pixel, shown in the right panel, is broad and highly skewed for every angular bin. This indicates that the \Ha \  emitting gas clouds associated with the jet is not produced by a smooth, uniformly filling halo. The mean values are possibly driven by a small number of bright, dense and highly ionized clumps embedded within a more diffuse, low surface brightness CGM component. This is consistent with our observations, and explains why we measure $>5\sigma$ detection in mean \Ha \ stacks along the radio axis, where some sightlines intersect these bright clumps. However, the azimuthally averaged signal and median-like statistics remain weak or undetected because the emitting gas has a low covering fraction. Note, however that the amplitudes or the absolute values of the \Ha\ surface brightness are lower by approximately three orders of magnitude compared to our observational data. This discrepancy indicates that photoionization by the metagalactic background alone is far too feeble to power the observed emission, and implies a substantial energetic boost driven by the impact of the radio source. 
The simulation therefore supports a multiphase picture in which radio jets propagate through an extended, clumpy cool CGM reservoir, similar to the broadly distributed Mg\,{\sc ii}-absorbing clouds observed in our DESI sightlines. Mg\,{\sc ii} absorption traces the presence of this cool gas over a wide range of azimuthal angles, whereas H$\alpha$ emission identifies the subset of clouds whose density, pressure, or ionization state has been enhanced by the jet or cocoon.  
Although this single SIMBA galaxy is an example rather than a one-to-one detailed model comparison to our sample, it demonstrates that jet-mode feedback can produce the qualitative signature inferred from the observations. 
Future comparisons with larger samples of simulated jet-mode galaxies, matched in stellar mass, radio power, halo mass, viewing angle, and jet age, will be essential for turning this qualitative agreement into quantitative constraints on how radio jets deposit energy into cool circumgalactic gas.

\subsection{Continuum subtraction of DESI quasar spectra and detailed stacking procedure}
\label{subsec:stacking}

We describe the details of continuum subtraction of our quasar sightline spectra from DESI, their stacking analyses and determining the noise baseline shown in Fig.~\ref{fig:angle_stack}. First, we search for \Ha\ line emission (rest wavelength $\lambda_0=6562.8$ \AA) at the redshift of the foreground radio galaxy by shifting each background-quasar spectrum into the foreground rest frame. 
We extract a local window of half-width $\Delta\lambda = 150~${\rm \AA} 
around the target line in the foreground rest frame. Within this window, we keep only pixels with finite flux and positive inverse variance.
Since the background sources are quasars themselves, their own broad or narrow emission features can fall into the same observed-frame spectral region as the foreground line of interest. We therefore mask common quasar rest-frame optical emission lines after projecting them into the radio-galaxy rest frame. For each sightline, a background-quasar line of rest wavelength $\lambda_{\rm QSO}$ appears in the foreground-radio-galaxy rest frame at
\begin{equation}
    \lambda_{\rm QSO\rightarrow RG}
    =
    \lambda_{\rm QSO}
    \frac{1+z_{\rm QSO}}{1+z_{\rm RG}} .
\end{equation}
We mask pixels within $\pm900~{\rm km~s^{-1}}$ of the projected positions of primarily strong nebular lines, particularly \oii $\lambda3727$, \hb, \oiii $\lambda\lambda4959,5007$, \NII $\lambda\lambda6548,6583$, \Ha, and \SII $\lambda\lambda6717,6731$, whenever those projected lines fall within the local stacking window. If the quasar-line mask would remove nearly all continuum pixels in a given spectrum, the mask is not applied to that object in order to prevent a small number of cases from eliminating otherwise usable spectra.

We subtract the local continuum $f_{{\rm cont},i}(\lambda)$ independently for each foreground--background pair. The continuum is modeled with a second-order polynomial in the foreground rest-frame wavelength after excluding the emission line region $|\lambda_{\rm rest}-\lambda_0| < 50~${\rm  \AA}. 
If too few continuum pixels remain for a polynomial fit, we subtract the median flux in the local window instead. The continuum-subtracted spectrum $f_{{\rm cs},i}(\lambda)$ is then
\begin{equation}
    f_{{\rm cs},i}(\lambda)
    =
    f_i(\lambda)-f_{{\rm cont},i}(\lambda).
\end{equation}


After continuum subtraction, each spectrum is interpolated onto a common rest-frame wavelength grid. The grid spacing is inferred from the median pixel spacing of the first spectrum with valid wavelength coverage in the target-line window. For each pixel in the common grid, the inverse-variance weight ($w_i(\lambda)$) is obtained by interpolating the original variance array and using
\begin{equation}
    w_i(\lambda)=\frac{1}{\sigma_i^2(\lambda)}.
\end{equation}
The fiducial stacked spectrum is the inverse-variance-weighted mean:
\begin{equation}
    \langle f_{\lambda}(\lambda)\rangle =
    \frac{\sum_i w_i(\lambda) f_{{\rm cs},i}(\lambda)}
    {\sum_i w_i(\lambda)}.
\end{equation}
We require each retained object to have at least five finite pixels in the full stacking window. For the fiducial H$\alpha$ stacks, we additionally require at least five finite pixels within $\pm50$ \AA\ of the target line and require at least five contributing spectra per wavelength pixel. Pixels failing the latter criterion are set to NaN. These cuts prevent noisy spectra and sparsely sampled wavelength regions from driving the result.

We measure the integrated H$\alpha$ flux by summing the continuum-subtracted stacked spectrum over a fixed velocity window centered on $\lambda_{\rm H\alpha}=6562.8$ \AA. We adopt $|\Delta v|<500~{\rm km~s^{-1}}$, corresponding to $\Delta\lambda=\lambda_{\rm H\alpha}\Delta v/c\simeq10.9$ \AA\ to avoid including the contribution of \NII \  in our measurements. The integrated flux is therefore
\[
F_{\rm H\alpha}
=
\int_{\lambda_{\rm H\alpha}-\Delta\lambda}^{\lambda_{\rm H\alpha}+\Delta\lambda}
\left[y(\lambda)-y_{\rm base}\right]d\lambda,
\]
where $y(\lambda)$ is the stacked flux density in units of $10^{-17}\ {\rm erg\ cm^{-2}\ s^{-1}\ \text{\AA}^{-1}}$ and $y_{\rm base}$ is the residual baseline estimated from adjacent line-free sidebands.

\vspace{5pt}
\subsection{Noise estimation and control stacks}
\label{subsec:noise_control}

We estimate the statistical uncertainty in two ways. First, for each stack we compute the empirical dispersion of the individual continuum-subtracted spectra contributing to each wavelength pixel. The standard error on the stacked spectrum is
\begin{equation}
    \sigma_{\rm stack}(\lambda)=
    \frac{{\rm std}\left[f_{{\rm cs},i}(\lambda)\right]}
    {\sqrt{N_{\rm contrib}(\lambda)}} ,
\end{equation}
where $N_{\rm contrib}(\lambda)$ is the number of spectra with finite continuum-subtracted flux at that pixel. This uncertainty captures object-to-object scatter, and wavelength-dependent variations in the number of contributing spectra.

Second, we construct blank-region noise stacks centered on randomly selected rest-frame wavelengths where no foreground H$\alpha$ or any other emission lines are expected. We used the same background quasar spectra, foreground redshifts, wavelength window, masking, continuum fitting, and inverse-variance stacking procedure to keep the entire analyses method consistent. We constructed several blank stacks centered at $6800$ \AA, $5500$ \AA\ and $4000$ \AA\ with the same $\pm150$ \AA\ window used during H$\alpha$ stacking. The blank stacks serve as a noise baseline for the H$\alpha$ measurement, as shown in Figure \ref{fig:angle_stack}. Since these are processed through the identical continuum-subtraction, quasar-line-masking, bad-pixel-cleaning, interpolation and analyses steps, they capture residual systematics in the most robust way possible. Hence in the main analyses, we evaluate whether the H$\alpha$ stacks show excess flux relative to this noise baseline.

\bibliography{main}
\bibliographystyle{aasjournal}
\end{document}